\documentstyle[aps,epsfig,float,amstex,axodraw]{revtex}

\newcommand{\be}{\begin{equation}}
\newcommand{\ee}{\end{equation}}
\newcommand{\beq}{\begin{eqnarray}}
\newcommand{\eeq}{\end{eqnarray}}

\newcommand{\cu}[0]{\mathrm{CuO}_{4} }

\newcommand{\ve}{\Vertex}

\restylefloat{figure}
\restylefloat{table}

\begin{document}
    
\def\gC{\mbox{\boldmath $C$}}
\def\gZ{\mbox{\boldmath $Z$}}
\def\gR{\mbox{\boldmath $R$}}
\def\gN{\mbox{\boldmath $N$}}
\def\ua{\uparrow}
\def\da{\downarrow}
\def\eq{\equiv}
\def\a{\alpha}
\def\b{\beta}
\def\g{\gamma}
\def\G{\Gamma}
\def\d{\delta}
\def\D{\Delta}
\def\e{\epsilon}
\def\ve{\varepsilon}
\def\z{\zeta}
\def\h{\eta}
\def\th{\theta}
\def\k{\kappa}
\def\l{\lambda}
\def\L{\Lambda}
\def\m{\mu}
\def\n{\nu}
\def\x{\xi}
\def\X{\Xi}
\def\p{\pi}
\def\P{\Pi}
\def\r{\rho}
\def\s{\sigma}
\def\S{\Sigma}
\def\t{\tau}
\def\f{\phi}
\def\vf{\varphi}
\def\F{\Phi}
\def\c{\chi}
\def\w{\omega}
\def\W{\Omega}
\def\Q{\Psi}
\def\q{\psi}
\def\de{\partial}
\def\inf{\infty}
\def\ra{\rightarrow}
\def\bra{\langle}
\def\ket{\rangle}

\title{
``Spin-Disentangled'' Exact Diagonalization  of Repulsive Hubbard 
Systems: Superconducting Pair Propagation}

\author{Michele Cini, Gianluca Stefanucci, Enrico Perfetto and Agnese Callegari}

\address{Istituto Nazionale di Fisica della Materia, Dipartimento di Fisica,\\
Universita' di Roma Tor Vergata, Via della Ricerca Scientifica, 1-00133\\
Roma, Italy}
\maketitle

\begin{abstract}     

By a novel exact diagonalization technique
we show that  bound pairs propagate between repulsive Hubbard 
clusters  in a superconducting fashion.  The size of the matrices that 
must be handled depends on the number of fermion configurations {\em 
per spin}, which is of the order of the square root of the overall size of the 
Hilbert space.
We use CuO$_{4}$ units  connected by weak O-O  
links  to model interplanar coupling and c-axis superconductivity
in Cuprates. The  numerical evidence on Cu$_{2}$O$_{8}$ and 
Cu$_{3}$O$_{12}$ prompts a new analytic scheme  describing  
the propagation of bound pairs and also the superconducting flux 
quantization in a 3-d geometry.
\end{abstract}
\vspace{1cm}

{\small 

Evidence for pairing in the repulsive Hubbard and related models has 
been reported by several authors.  Analytic approaches 
\cite{EPJB2000}\cite{SSC1999}, even at strong coupling\cite{citro}, 
generalized conserving approximation theories like FLEX\cite{flex}, 
as well as  Quantum Monte  Carlo Studies on supercells\cite{fettes} 
lead to this conclusion. However, we want more evidence about the real 
nature of the pairing interactions. Su and coworkers\cite{su} 
have reported that in narrow bands one- and two-dimensional  
Hubbard models no kind of superconducting long-range order holds at any 
non-zero temperature. Here we wish to explore the possibility that the 
Hubbard model can show superconductivity in the ground state when interplanar 
coupling is allowed. Since one cannot master the problem with an infinite stack 
of infinite planes, some economy is needed. However in high-$T_{c}$ superconductors 
the coherent length is $\sim$ a few lattice constants, and Cu-O planes can be approximately 
represented by clusters that are large enough to host a bound pair.   
 The {\em interplanar} hopping does not dissolve  pairs and
 superconducting flux quantization is their clear signature. 
The magnetic properties of {\em attractive} Hubbard models have been 
studied   by Canright and Girvin\cite{canright}; here we propose a 
 gedankenexperiment very much in the spirit of Little and 
Parks\cite{lipa},  in the {\em repulsive} case.
 
The repulsive Hubbard Hamiltonian of fully symmetric clusters 
${\cal C}$ has two-body singlet eigenstates without double 
occupation\cite{cibal1}\cite{cibal2}\cite{cibal3}\cite{cibal4} called 
$W=0$ pairs. The presence of such solutions at the highest occupied 
level of the non-interacting (Hubbard $U \rightarrow 0$) system is necessary to 
allow $\D_{{\cal C}}(N)<0$ where $\D_{{\cal C}}(N)=E^{(0)}_{{\cal C}}(N)+
E^{(0)}_{{\cal C}}(N-2)-2E^{(0)}_{{\cal C}}(N-1)$, and $E^{(0)}_{{\cal C}}(N)$ 
is  the interacting ground state energy of  the cluster ${\cal C}$ 
with $N$ fermions. By means of a non-perturbative canonical 
transformation\cite{SSC1999}\cite{cbs99}, it can also be shown that 
$\D_{{\cal C}}(N)<0$ is due to an attractive pairing effective interaction and 
at weak coupling  $|\D_{{\cal C}}(N)|$ is just the 
binding energy of the pair. 

$\cu$ is the smallest cluster which fully preserves the point-symmetry of 
the Copper-Oxide planes of high-$T_{c}$ materials. We have already described 
$W=0$ pairing  in great detail as a function of the one-body and interaction 
parameters on all sites; the study was extended to larger clusters 
too\cite{EPJB2000}\cite{cibal3}. $W=0$ bound pairs in the $\cu$ 
cluster are found to exist in the physical region of the parameter space. 
However, since it is the symmetry that produces the pairing 
force we  use the simplest working model to study bound pair 
propagation. Here, in order to 
simplify the analytical formulas,  
we neglect the O-O hopping term and also any distinction between Cu and 
O sites (except geometry, of course). The only nonvanishing hopping matrix 
elements are those between an Oxygen site and the central Copper site; they 
are all equal to $t$. For the sake of simplicity, we parametrize 
the Hubbard model in such a way that actually everything depends only on the 
ratio $U/t$; the important thing is that in this way we still have 
access to the part of the parameter space where pairing 
occours\cite{cibal1}. Thus, we  consider the Hubbard Hamiltonian 
\begin{equation}
H_{\mathrm{CuO}_{4}}=t \sum_{i\s}( d^{\dag}_{\s}p_{i\s}+p_{i\s}^{\dag}d_{\s})+
U(\sum_{i}\hat{n}^{(p)}_{i\ua}\hat{n}^{(p)}_{i\da}+
\hat{n}^{(d)}_{\ua}\hat{n}^{(d)}_{\da})
\label{cuo4ham}
\end{equation}
where $p^{\dag}_{i\s}$ and $p_{i\s}$ are the  creation and 
annihilation operators onto the Oxygen $i=1,..,4$ with spin 
$\s=\ua,\da$, $d^{\dag}_{\s}$ and $d_{\s}$ are the  creation and 
annihilation operators onto the Copper site,  
while $\hat{n}^{(p)}_{i\s}=p^{\dag}_{i\s}p_{i\s}$ and 
$\hat{n}^{(d)}_{\s}=d^{\dag}_{\s}d_{\s}$ are the 
corresponding number operators. $H_{\mathrm{CuO}_{4}}$ is invariant 
under the permutation Group $S[4]$, which has the irreducible 
representations ({\em irreps}) ${\cal A}_{1}$ (total-symmetric), 
${\cal B}_{2}$ (total-antisymmetric), ${\cal E}$ (self-dual), 
${\cal T}_{1}$ and its dual ${\cal T}_{2}$, of dimensions 
1, 1, 2, 3 and 3, respectively.  The ground state of $\cu[2]$  
(i.e. $\cu$ with 2 fermions) belongs to $^{1}{\cal A}_{1}$ and that of 
$\cu[4]$ is in $^{1}{\cal E}$; both are singlets, as the notation 
implies.  
The ground state of $\cu[3]$ is a $^{2}{\cal T}_{1}$ doublet. 
$\Delta_{\cu}(4)<0$ for this model when $0< U \lesssim 34.77\;t$, as 
shown in Fig.(\ref{delta}).
\begin{figure}[H]
\begin{center}
	\epsfig{figure=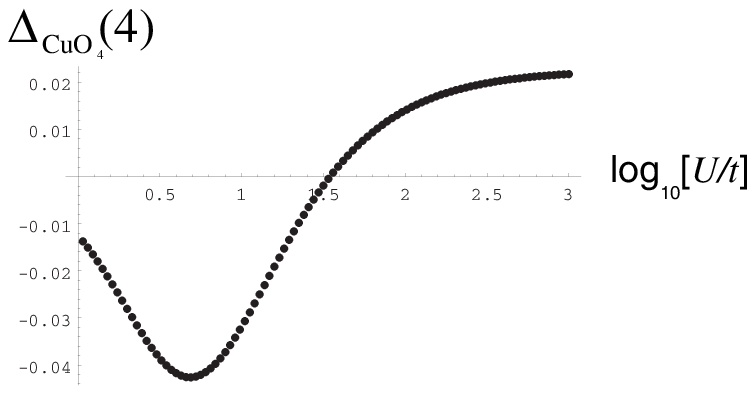,width=5cm}
	\caption{\footnotesize{Trend of $\D_{\cu}(4)$ in $t$ units versus $\log[U/t]$ .}}
    \label{delta}
\end{center} 
\end{figure}
Thus, we introduce a graph $\L$ with CuO$_{4}$ units as nodes.
The total Hamiltonian is 
\begin{equation} 
H_{\rm tot} = H_{0}+H_{\t}.
\label{hlattice}
\end{equation}
with 
\begin{equation}
H_{0}=\sum_{\a\in \L} \left[ t \sum_{i\s}( d^{\dag}_{\a \s}p_{\a, i\s}+
p_{\a, i\s}^{\dag}d_{\a\s})+
U(\sum_{i}\hat{n}^{(p)}_{\a, i \ua}\hat{n}^{(p)}_{\a, i\da}+
\hat{n}^{(d)}_{\a\ua}\hat{n}^{(d)}_{\a\da})  \right] \; ,   
\label{senzatau} 
\end{equation}
where $p^{\dag}_{\a, i\s}$ is the  creation operator 
onto the Oxygen $i=1,..,4$ of the $\a$-th cell and so on. Hence, 
the point symmetry Group of $H_{0}$ is $S[4]^{|\L|}$, with $|\L|$  the number of nodes.
There are many different ways to model an {\it inter-planar} hopping.
Nevertheless, to preserve the symmetry that produces 
the $\D_{\mathrm{CuO}_{4}}(4)<0$ property,  $H_{\tau}$ must be invariant 
under the $S[4]$ diagonal subgroup of $S[4]^{|\L|}$. In the following we shall consider a 
hopping term allowing to a particle in the $i$-th Oxygen site of the 
$\a$-th unit to move towards the  $i$-th Oxygen site of the 
$\b$-th unit with hopping integral $\t_{\a\b}$:
\begin{equation}
H_{\tau}=\sum_{\a,\b\in\L}\sum_{i\s}
\left[\t_{\a\b} p_{\a,i\s}^{\dag}p_{\b,i\s}+{\mathrm h.c.}\right]\; .
\label{htau}
\end{equation} 

For $N=2|\L|$ and $\tau_{\a\b}\equiv 0$, the unique  ground state consists of 2 fermions in
each CuO$_{4}$ unit. This paper is devoted to  the inter-planar hopping 
produced by  small $\tau_{\a\b} \ll |\Delta_{\cu}(4)|$ with a  total number of particles
$N=2|\L|+2p$;  $p$ represents the number of added pairs. When $U/t$ is such that 
$\D_{\mathrm{CuO}_{4}}(4)<0$, each pair prefers to lie on a single 
$\cu$  and for $N=2| \L | + 2p$ the unperturbed ground state is 
$2^{p}$$\times$${|\L|}\choose{p}$ times degenerate 
(since $^{1}{\cal E}$ has dimension 2).

By this sort of models one can study the interaction of several fermion 
pairs in the same system. The simplest topologically non-trivial graph is 
the ring, with a set $\L=\{1,2,\ldots,|\L|\}$ and 
\begin{equation}
\t_{\a\b}=\left\{\begin{array}{ll}
\t & {\mathrm if}\;\; \b=\a+1, \; \\
\t^{\ast} & {\mathrm if}\;\; \b=\a-1, \; \\
0 & {\mathrm otherwise}\;.\end{array}\right.\;\quad\quad
\;\t=|\t|e^{\frac{2\pi i}{|\L|}\frac{\f}{\f_{0}}}.
\label{tau}
\end{equation}
where $\f$ is the magnetic flux concatenated by the ring and 
$\f_{0}=\frac{h c}{e}$. In the absence of magnetic field, $\t$ will be taken to be real.  
    
Note that for $p$=0 the concentration (number of holes per atom) 
is $2/5=0.4$; this is somewhat more than half-filling 
($1/3\approx 0.33$) but still reasonable. 
We are using  CuO$_{4}$ as the unit just for the sake of 
simplicity, but  the $W=0$ mechanism produces bound pairs at 
different fillings for larger clusters\cite{cibal4} and the full plane\cite{EPJB2000}\cite{SSC1999} 
too. By replacing CuO$_{4}$ by larger units one can  model other ranges of the hole concentration. 

We   exactly diagonalize the $|\L|=2$ and $|\L|=3$ ring Hamiltonian;  
to this end we introduce the {\em Spin-Disentangled}  technique.
We let  $M_{\ua}+M_{\da}=N$ where $M_{\s}$ is the number of  particles 
of spin $\s$; $\{|\f_{\a\s}\ket\}$ is a real orthonormal basis, that 
is, each vector is a homogeneous polynomial in the $p^{\dag}$ and $d^{\dag}$ 
of degree $M_{\s}$ acting on the vacuum. We write the 
ground state wave function in the form 
\begin{equation}
|\Psi\rangle=\sum_{\alpha \beta}
L_{\alpha \beta}|\phi_{\alpha \uparrow}\rangle 
\otimes |\phi_{\beta \downarrow} \rangle 
\label{lali}
\end{equation}
which shows how the $\ua$ and $\da$ configurations are entangled. The 
electrons of one spin are treated as the ``bath'' for those of the 
opposite spin: this form also enters the proof of a famous theorem by Lieb\cite{lieb}.
In Eq.(\ref{lali})   $L_{\a\b}$ is a 
$m_{\ua}\times m_{\da}$ rectangular matrix with 
$m_{\s}$=${5|\L|}\choose{M_{\s}}$. 
We let  $K_{\s}$ denote the kinetic energy $m_{\s}\times m_{\s}$ 
square matrix of $H_{\rm tot}$ in the basis $\{|\f_{\a\s}\ket\}$, 
and $N^{(\s)}_s$ the spin-$\s$ occupation number matrix at site 
$s$ in the same basis ($N^{(\s)}_s$ is a  symmetric matrix since the 
$|\f_{\a\s}\ket$'s are real). Then, $L$ is acted upon by the Hamiltonian 
$H_{\rm tot}$ according to the rule 
\begin{equation}
H_{\rm tot}[L]=[K_{\ua}L + L K_{\da}]+
U \sum_{s}  N^{(\ua)}_s L N^{(\da)}_s\;.
\label{lieb3}
\end{equation}
In particular for $M_{\ua}=M_{\da}$ ($S_{z}=0$ sector) 
it holds $K_{\ua}=K_{\da}$ and $N^{(\ua)}_s=N^{(\da)}_s$. Thus, the action 
of $H$ is obtained in a spin-disentangled way.
In the $S_{z}=0$ sector for $|\L|$=3 the size of the problem is 
1863225 and the storage of the Hamiltonian matrix requires  much space; 
by this device, we can work with matrices whose dimensions is the 
square root of those of the Hilbert space: $1365 \times 1365$  matrices 
solve the $1863225 \times 1863225$ problem, and are not even required to be 
sparse. We believe that this approach will be generally useful for 
the many-fermion problem. Since we are mainly interested in getting the 
low-lying part of the spectrum as fast as possible we opted for the Lanczos method,
taking advantage from 
our knowledge of the $S[4]$ irrep of the $\tau=0$ ground state; the 
scalar product is given by $\bra \Q_{1}|\Q_{2} \ket = 
{\mathrm Tr}(L_{1}^{\dag}L_{2})$. In this way, 
the Hamiltonian matrix takes the tri-diagonal form; however a 
numerical instability sets in well before convergence is achieved  if 
one uses chains longer than a few tens of sites. Therefore we use 
repeated two-site chains alternated with moderate-size ones. 

The two-$\cu$ ring (14,400 configurations in the $S_{z}=0$ sector) is readily solved
by a Mathematica code on the personal computer; however, this cluster 
is not adequate for studying the quantization 
(superconducting or otherwise) of a magnetic flux by the bound pair. 
The reason is that the two units are each at the left {\em and} at the 
right of each other; any vector potential perpendicular to the 
$\cu$'s can always be gauged away. However, we have 
verified that the ground state energy  with  6 holes is 
$E_{\mathrm{Cu}_{2}\mathrm{O}_{8}}^{(0)}(6)=
E_{\cu}^{(0)}(4)+E_{\cu}^{(0)}(2)$  for $\tau=0$ and it receives a   
negative correction $\propto \tau^{2}/|\Delta_{\cu}(4)|$ for small $\t$, 
which is consistent with the presence of a bound pair. 
\begin{figure}[H]
\begin{center}
	\epsfig{figure=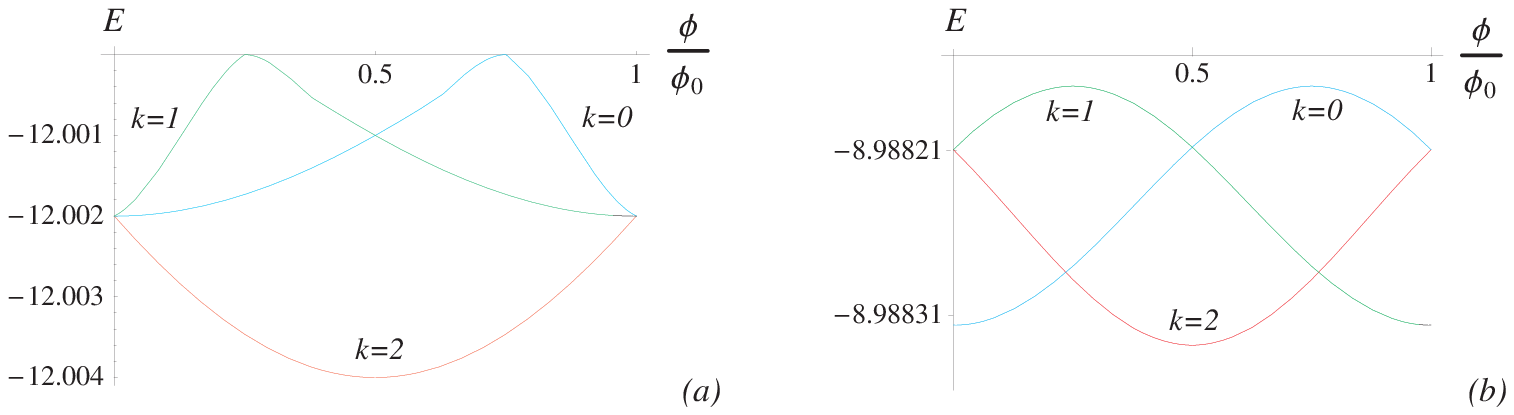,width=13cm}
	\caption{\footnotesize{Numerical results for Cu$_{3}$O$_{12}$  with 
	 $\tau=0.001$. Lowest-energy eigenvalues  labeled by their 
	interplanar quasi-momentum are shown versus flux $\phi$. The pattern 
	is periodic (a flux quantum can be gauged away). $(a)$ $U=0$.
	A paramagnetic current is excited by the field and the system is 
	utterly normal. $(b)$ $U=5$.  The ground state shows a clear superconducting 
	pattern, with a minimum at $\phi = \frac{\phi_{0}}{2}$. All energies are 
	in $t$ units.}}
    \label{u0u5}
\end{center} 
\end{figure}
The three-$\cu$ ring behaves similarly, but can also concatenate 
a flux. In Figure (\ref{u0u5}.$a$) and (\ref{u0u5}.$b$) we show the lowest 
eigenvalues versus $\phi$ for $U=0$ and $U=5\; t$, respectively; $k$ denotes 
the interplanar quasi-momentum quantum label. At $\tau=0$ the ground state 
energy is $E_{\mathrm{Cu}_{3}\mathrm{O}_{12}}^{(0)}(8)=
E_{\cu}^{(0)}(4)+2 E_{\cu}^{(0)}(2)$  and the low energy 
sector derives mainly from the tensor product of the ground states of 
three independent $\cu$'s with 4, 2 and 2 holes (which is the 
foundamental multiplet) and 3, 3 and 2 holes (which is the lowest lying 
excited multiplet separated by a gap $\Delta_{\cu}(4)$).  
For $U=0$, see Fig.(\ref{u0u5}.$a$), there is no pairing in $\cu$ and 
indeed the ground state energy is linear in the field at small fields 
(normal Zeeman effect). The lowest state is $k=2$ throughout. Interestingly,  
Cu$_{3}$O$_{12}$ {\em concatenated with half a flux quantum} would be diamagnetic, 
but the absence of a second minimum shows that it would be Larmor diamagnetism.
By contrast, at $U$=5 $t$, when pairing in $\cu$  is about optimum, see 
Fig.(\ref{delta}), the $k=2$ state 
is lowest in the central sector, $k=0$ is the ground state at $\phi \rightarrow 0$ 
while $k=1$ is lowest as $\phi \rightarrow \phi_{0}$, see Fig.(\ref{u0u5}.$b$); 
this produces level crossings and the superconducting flux quantization; there 
is a central minimum when the system swallows a half quantum of flux while, as we 
verified, $\Delta_{\mathrm{Cu}_{3}\mathrm{O}_{12}}< 0$. Remarkably, one also observes 
superconducting quantization of a magnetic flux orthogonal to the 
plane\cite{cibal4}. With increasing $U/t$, the binding energy of the pair starts 
decreasing and eventually vanishes for $U \approx 34.77\;t$; we have found 
that at this point the flux quantization returns normal and the 
system behaves like a paramagnet. Even at optimal $U$,  the side 
barriers are depressed by increasing $\tau$; at $\tau \sim 0.1 \; t$ only 
small remnants remain; for still larger interplanar hoppings the 
superconductivity is removed and a pattern similar to 
Fig.(\ref{u0u5}.$a$) prevails.  For larger $|\L|$,  a smaller supercurrent 
would be necessary to screen the half flux quantum and the critical $\tau$ 
which kills the superconductivity 
should  be expected to grow larger.

In order to better analyze the results physically and extend them
qualitatively to arbitrary graphs, we obtained an effective Hamiltonian
by the cell-perturbation method with $H_{0}$, Eq.(\ref{senzatau}), the
``cell-Hamiltonian'' and $H_{\t}$, Eq.(\ref{htau}), the
``inter-cell perturbation'' and by taking into account only the
low-energy
singlet sector. We note that the cell-perturbation method was already
used in Ref.\cite{jefferson} to support the original Anderson's
conjecture\cite{anderson} on the ``low-energy equivalence'' between the
$d-p$ model
(proposed by Emery\cite{emery}) and the single-band Hubbard model.
Despite the analogies with Ref.\cite{jefferson} (like the same
cell-Hamiltonian
and weak O-O links between different cells) our inter-cell
perturbation is different and, more important, it is the
low-energy sector which differs (one needs to consider $\cu$ units with 2, 3 and 4
holes to get bound pairs, in contrast with 0, 1 and 2 holes of
Ref.\cite{jefferson}).

For a general graph $\L$, with $2|\L|+2p$ holes, we treat $H_{\tau}$ by a 
simplified second-order degenerate perturbation 
theory, since  $H_{\t}$ is a one-body operator. Each 
degenerate unperturbed ground state  $|\F_{0}^{{\cal S}}\ket$ may be 
labelled by a set  ${\cal S}\subset\L$  of 
units occupied by four holes;  $|{\cal S}|=p$. The secular problem 
yields the  eigenvalue equation 
\begin{equation}
\frac{1}{\D_{\cu}(4)}\sum_{q}\sum_{{\cal S}'}{\bra\F^{{\cal S}}_{0}|H_{\t}|\F_{q}\ket
\bra\F_{q}|H_{\t}|\F^{{\cal S}'}_{0}\ket}a_{{\cal S}'}=\ve a_{{\cal S}}\;
\label{schr1}
\end{equation}
where the sum has been truncated to the low-energy excited eigenstates 
involving $\cu$ units with $2\leq n \leq 4$  holes, all taken in their ground 
states $|\Psi^{(n)}_{0}(\a)\ket,\;\a=1\ldots |\L| $. 
The amplitude $a_{{\cal S}}\equiv a(\a_{1},\ldots,\a_{p})$ is totally symmetric 
with respect to permutations of the distinct indices $\a_{1},\ldots,\a_{p}$. Letting  
${\cal C}(\a)=\{\b\in\L:\t_{\a\b}\neq 0\}$, after some algebra 
Eq.(\ref{schr1}) may be written in the form:
\begin{equation}
\ve a(\a_{1},\ldots,\a_{p})=\sum_{j=1}^{p}
\sum_{\b\in{\cal C}(\a_{j})} \left[{\cal T}^{^{\rm Bose}}_{\b,\a_{j}}
a(\a_{1},..,\a_{j-1},\b,\a_{j+1},..,\a_{p})-
|{\cal T}^{^{\rm Bose}}_{\b,\a_{j}}| 
\prod_{i\neq j}(1-\d_{\b\a_{i}})
a(\a_{1},\ldots,\a_{p})\right].
\label{schrexpl}
\end{equation} 
This is a Schr\"odinger equation for $p$ hard-core bosons hopping 
with an effective hopping integral ${\cal T}^{^{\rm Bose}}_{\a,\b}\equiv (\t_{\a\b}^{{\mathrm 
eff}})^{2}/\D_{\cu}(4)$, with  
\begin{equation}
\t_{\a\b}^{{\mathrm eff}}=
\bra\Psi^{(2)}_{0}(\a)|\otimes\bra\Psi^{(4)}_{0}(\b)|H_{\t}
|\Psi^{(3)}_{0}(\a)\ket\otimes
|\Psi^{(3)}_{0}(\b)\ket.
\end{equation}
In Fig.(\ref{tauflu}.$a$) we show the 
trend of $(|\t_{\a\b}^{{\mathrm eff}}|/|\t_{\a\b}|)^{2}$ versus $U/t$; we 
note that the ratio decreases monotonically.
 In Eq.(\ref{schrexpl}), the first  term in the r.h.s. describes hole pair 
 propagation, {\em e.g.} from unit $\a_{j}$ to an unoccupied unit $\b$; in 
 the second sum, the system gets back to the initial state after 
 virtually exploring unit $\b$;  the term $\prod_{i\neq j}(1-\d_{\b\a_{i}})$ 
takes into account that if  $\b$ is  one of the 
occupied units, the 
particle cannot move toward it. Due to the minus sign, the term in $
|{\cal T}^{^{\rm Bose}}_{\b,\a_{j}}| $
represents pair-pair repulsion.
\begin{figure}[H]
\begin{center}
	\epsfig{figure=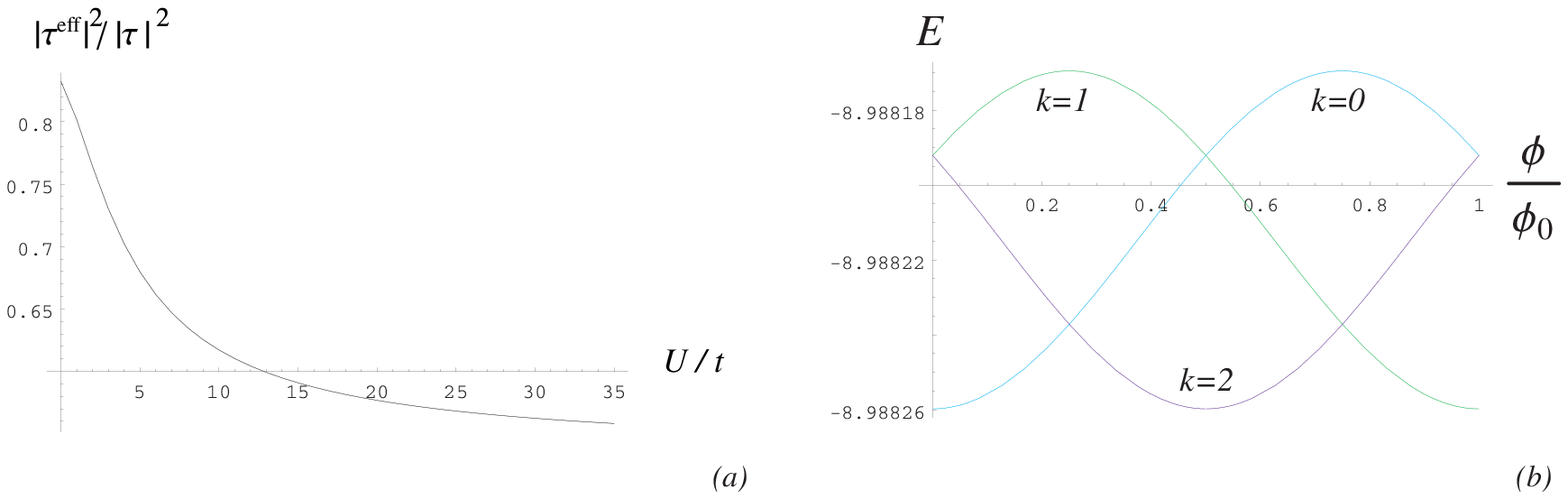,width=12cm}
	\caption{\footnotesize{$(a)$ $(|\t_{\a\b}^{{\mathrm 
	eff}}|/|\t_{\a\b}|)^{2}$ versus $U/t$. $(b)$
	Results  of Eq.(\ref{schrexpl}) for Cu$_{3}$O$_{12}$  with 
	 $\tau=0.001\; t$, $U=5\; t$. Lowest-energy eigenvalues  labeled by their 
	interplanar quasi-momentum are shown versus flux $\phi$. All energies are in $t$ 
	units. }}
    \label{tauflu}
\end{center} 
\end{figure}
In Fig.(\ref{tauflu}.$b$) the superconducting flux-quantization for 
the $|\L|=3$ ring is reported as reproduced by solving Eq.(\ref{schrexpl});
it agrees well both qualitatively and quantitatively with the numerical results of 
Fig.(\ref{u0u5}.$b$), thus confirming the above approximation. More data and a fuller 
account of the low-energy theory will be presented elsewhere.

In conclusion, we used a set of   CuO$_{4}$ units  connected by weak O-O  
links  to model interplanar coupling and c-axis superconductivity
in Cuprates. The results show that the system 
with two holes in each  unit is a background such that inserting  
$2p$ holes  one gets $p$ pairs,  bound by the repulsive interaction. The bound 
pairs propagation is well described by Eq.(\ref{schrexpl}). We found 
analytically the superconducting flux quantization in the ring-shaped 
systems and confirmed this finding numerically for  the 3-unit ring
(1,863,225 configurations). To this end, we  introduced a novel exact-diagonalization 
technique, which reduces the size of the matrices that 
must be handled to the square root of the overall size of the 
Hilbert space.  Actually, real systems contain also vertical 
links via the orbitals of the apical oxygens. We expect that the inclusion 
of these hoppings does not change qualitatively the results since   
they do not contribute to the propagation of the bound pair in the 
lowest-order approximation.

\begin{center}
{\bf Acknowledgement}
\end{center}
We thank Dr. Claudio Verdozzi for useful comments.

}

\begin{center}
\end{center}

\end{document}